\documentclass[twocolumn,showpacs,preprintnumbers,amsmath,amssymb,superscriptaddress]{revtex4}

\usepackage{amsmath,amssymb,enumerate,bbm,calc,capt-of,ifthen}
\usepackage{graphicx}
\usepackage{dcolumn}
\usepackage{bm}
\usepackage{epsfig}

\newtheorem{cntr}{Counter DO NOT USE}

\newtheorem{varremark}[cntr]{Remark}

\newenvironment{proof*}[1]{
  \noindent\textbf{#1\ }}{\hspace*{\fill}
  \begin{math}\Box\end{math}\medskip}

\newcommand{\tmfloatcontents}{}
\newlength{\tmfloatwidth}
\newcommand{\tmfloat}[5]{
  \renewcommand{\tmfloatcontents}{#4}
  \setlength{\tmfloatwidth}{\widthof{\tmfloatcontents}+1in}
  \ifthenelse{\equal{#2}{small}}
    {\ifthenelse{\lengthtest{\tmfloatwidth > \linewidth}}
      {\setlength{\tmfloatwidth}{\linewidth}}{}}
    {\setlength{\tmfloatwidth}{\linewidth}}  \begin{minipage}[#1]{\tmfloatwidth}
    \begin{center}
      \tmfloatcontents
      \captionof{#3}{#5}
    \end{center}
  \end{minipage}}

\newcommand{\abs}[1]{\left| #1 \right|}

\newcommand{\Mo}[0]{M_0}

\numberwithin{cntr}{section}

\begin{document}

\title{Exact Results for Ionization of Model Atomic Systems}
\author{O.Costin}
\affiliation{Department of Mathematics, The Ohio State University, Columbus, Ohio}
\author{J.L. Lebowitz}
\affiliation{Department of Mathematics and Physics, Rutgers University, Piscataway, New Jersey}
\author{C. Stucchio}
\affiliation{Department of Mathematics, Rutgers University, Piscataway, New Jersey}
\author{S. Tanveer}
\affiliation{Department of Mathematics, The Ohio State University, Columbus, Ohio}

\date{\today}

\begin{abstract}
  We present rigorous results for quantum systems with both bound and continuum states subjected to an arbitrary strength time-periodic field. We prove that the wave function takes the form $\psi(x,t) = \sum_{j} \sum_{k=0}^{N_{j}-1} t^{k} e^{-i \sigma_{j} t}c_{j} \phi_{j,k}(x,t) + \psi_{d}(x,t)$, with $\phi_{j,k}(x,t)$ a set of time-periodic resonant states with quasi-energies $\sigma_{j}=E_{j}-i\Gamma_{j}/2$, with $E_{j}$ the Stark-shifted energy and $\Gamma_{j}$ the ionization rate, and $N_{j}$ the multiplicity of each resonance. $\psi_{d}(x,t)$ is the dispersive part of the the solution, and is given by a power series in $t^{-1/2}$. Generally, $\Gamma_{j} > 0$ for each $j$ leading to ionization of the atom, but we also give examples where $\Gamma_{j}=0$ and implying the existence of a time-periodic Floquet bound state. The quantity $\sigma_{j}=E_{j}-i\Gamma_{j}/2$ has a convergent perturbation expansion for small field strengths. 
\end{abstract}

\maketitle

The ionization of atoms subjected to external time-dependent fields is an issue of central importance in atomic physics. There exist a variety of methods for treating this problem, including perturbation theory (Fermi's golden rule), numerical integration of the time-dependent Schr\"odinger equation, semi-classical phase-space analysis, Floquet theory and Complex dilations \cite{susskind:multiphoton,cohen:atomPhotonInteractions,solovev:ionizaitionRecombination,solovev:ionizaitionOfHydrogen,yajima:ACStarkExteriorScaling,galtbayer:floquetalternative,yajima:ACStarkResonances}. Still, there are very few rigorous results proving or disproving ionization by a periodic field of arbitrary strength and frequency, even for the simplest systems with both bound and continuum states. Such results are clearly desirable from both a theoretical and practical point of view. Numerical results are also difficult since delocalization of the wavefunction of the electron creates truncation errors at the boundary of the computational domain (absorbing potentials can help, but may also cause difficult to detect errors \cite[Section 3.2]{us:TDPSFjcp}). This has motivated us to undertake a systematic study of this problem in the context of non-relativistic quantum mechanics where the radiation field causing the ionization is treated classically\cite{cohen:atomPhotonInteractions}. We present here rigorous results applicable to the behavior of such systems in very intense laser fields, a subject of much current interest \cite{solovev:ionizaitionRecombination,solovev:ionizaitionOfHydrogen}.

For weak fields our results agree with those of perturbation theory. They show, in fact, that certain quantities such as the ionization rate and Stark shift have convergent expansions in the field strength, thus rigorously justifying the use of perturbation theory for weak fields. For larger fields, however, the behavior is very complex even in the simplest model systems and the results can be qualitatively different from those given by perturbation theory. In particular the ionization probability may not be monotone in the strength of the field, leading to some kind of stabilization \cite{costin:coulomb,costin:jpaionize,us:dipoledelta,MR1618647,schrader:anastab}.

In our work, we study the long time behavior of the solution of the Schr\"odinger equation in $d$ dimensions
\begin{equation}
  \label{eq:schro}
  i \partial_{t}\psi(x,t) = \left[-\frac{1}{2}\Delta + V_0(x) + V_{1}(x,t)\right] \psi(x,t)
\end{equation}
The units are chosen so that $\hbar = m = 1$. Here, $x \in \mathbb{R}^{d}$, $t \geq 0$, $V_0(x)$ is a binding potential having bound and continuum states, and $V_{1}(x,t) = \sum_{j=1}^{\infty} \left[\Omega_{j}(x) e^{i j \omega t} +c.c. \right]$ is a time-periodic potential representing the radiation field. Some of our results require that $\Omega_{j}(x)=0$ for $j > M \geq 1$. The initial condition $\psi(x,0)=\psi_{0}(x)$ is taken to be localized and of unit norm; for convenience we also assume $\psi_{0}(x)=0$ for $\abs{x} > R$ with $R$ arbitrary. 

Of primary interest is whether the system ionizes under the influence of the forcing $V_{1}(x,t)$, as well as the rate of ionization if it occurs. Ionization corresponds to delocalization of the wavefunction as $t \rightarrow \infty$. In particular we say that the system, e.g. an atom, will fully ionize if the probability of finding the particle in any bounded spatial region $B \subset \mathbb{R}^{d}$ goes to zero as time becomes large, i.e. $\int_{B} \abs{\psi(x,t)}^{2} dx \rightarrow 0$ as $t \rightarrow \infty$. Our results however go beyond this and give in many cases a comprehensive qualitative picture of the time evolution of $\psi(x,t)$. 

The time evolution is described by a resonance expansions for $\psi(x,t)$. That is, under the simplifying assumption that $\psi(x,0)$ is compactly supported, we obtain the decomposition of $\psi(x,t)$
\begin{equation}
  \label{eq:resonanceExpansion}
  \psi(x,t) = \sum_{j=0}^{M} \sum_{k=0}^{N_{j}-1} t^{k} e^{-i \sigma_{j} t}c_{j} \phi_{j,k}(x,t) + \psi_{d}(x,t)
\end{equation}
where $M$ is the number of bound states and resonances (possibly infinite), and $N_{j}$ is the multiplicity of the resonance $\sigma_{j}$. The exponents $\sigma_{j}=E_{j}-i \Gamma_{j}/2$ are complex numbers, with $E_{j}= \Re \sigma_{j}$ interpreted as a Stark shifted energy and $\Gamma_{j}=-2 \Im \sigma_{j}$ the ionization rate. The functions $\phi_{j,k}(x,t)$ are time-periodic Gamow vectors \cite{gamow:first} with period $2\pi/\omega$. This means the $\phi_{j,k}(x,t)$ are smooth eigenvalues of the Floquet Hamiltonian,
\begin{equation}
  \label{eq:floquet}
  \left[\left( -i \partial_{t} -\frac{1}{2}\Delta + V_0(x)+V_{1}(x,t)\right) -  \sigma_{j} \right]\phi_{j,k} = \phi_{j,k-1}
\end{equation}
where $\phi_{j,-1}=0$. Eq. \eqref{eq:floquet} is supplemented with an ``outgoing wave'' boundary condition which depends on the potentials $V_{0}(x)$ and $V_{1}(x,t)$. When $\Im \sigma_{j} \leq 0$, this boundary condition causes $\phi_{j,k}(x,t)$ to be exponentially growing in space. On the other hand, if $\Im \sigma_{j} = 0$, then $\phi_{j,k}(x,t)=0$ for $k \geq 1$ and $\phi_{j,0}(x,t)$ is a standard $L^{2}$ eigenvalue of the Floquet Hamiltonian. Our analysis can be used to make rigorous calculations involving these Gamow vectors,  e.g. \cite{gamow:first,ho:012102}.

In the case when $V_{0}(x)+V_{1}(x,t)=0$ for $\abs{x} > R$ in three dimensions, then $\phi_{j,0}(x,t)$ takes the form
\begin{multline}
  \phi_{j,0}(x,t) = \sum_{k} e^{-i k \omega t}\\
  \sum_{l=0}^{\infty} 
  \left[
    \sum_{m=-l}^{l} \phi_{j}^{l,m} Y_{l}^{m}(\theta,\varphi)
  \right]
  h^{(1)}_{l}(\abs{x} \sqrt{-\sigma_{j}\!\!-k\omega} )
\end{multline}
for $\abs{x} \geq R$. The square root is chosen to make the smaller angle with the real axis, and $h^{(1)}_{l}(z)$ is the spherical Bessel function of the third kind.

The last term, $\psi_{d}(x,t)$ is the dispersive part of the solution and has the expansion
\begin{equation}
  \label{eq:dispersiveExpansion}
  \psi_{d}(x,t) \sim \sum_{k=1}^{\infty} d_{k}(x,t) t^{-k/2}
\end{equation}
Generically $d_{1}(x,t)=0$, but the presence of a zero energy resonance causes $d_{1}(x,t)$ to be nonzero. In the case of finite range $V_{0}(x)$ and $V_{1}(x,t)=0$, we can resum this series using the Borel summability procedure \cite{MR1625999,MR1856397}, and we conjecture that this can be done in general \cite{costin:jspionize}. It follows from \eqref{eq:dispersiveExpansion} that for very large times the decay of any initially localized states will be given by a power law, which is expected since solutions to the Schr\"odinger equation with $V_{0}(x)=V_{1}(x,t)=0$ behave this way as well. In the cases studied the power law decay is dominant only for times when the survival probability in any bound state is already very small, but it has in fact been observed experimentally \cite{rothe:163601}.

Equation \eqref{eq:resonanceExpansion} provides a rigorous definition of the ionization rate and Stark shifted energy regardless of the strength of the field. Even in the case when $V_{1}(x,t)=0$, the Hamiltonian $H_{0}=-(1/2)\Delta+V_{0}(x)$ will have many resonances \cite{MR1668841} (complex energies $E_{j}-i\Gamma_{j}/2$ for which Gamow vectors can be found). As $V_{1}(x,t)$ is ``switched on'', the bound states will (generically) become resonances, while the already existing resonances will change as well. In general, letting the parameter $\epsilon$ represent the strength of the field, the quasi-energies $\sigma_{j}(\epsilon)$ will vary analytically in $\epsilon$, except for an isolated set of points where $\sigma_{j}(\epsilon)$ may have singularities. The bound states and resonances change analytically as well, and bound states become resonances when $\Gamma_{j}(\epsilon)=2 \Im \sigma_{j}(\epsilon)$ becomes nonzero.  

When the initial state is not compactly supported there may be additional exponentially or polynomially decaying terms which depend explicitly on $\psi_{0}(x)$. These terms are not resonances, and are present even in the solution of the free Schr\"odinger equation ($V_{0}(x)=V_{1}(x,t)=0$), and can be distinguished from resonances by this fact.

Given the expansion \eqref{eq:resonanceExpansion}, it follows that the only way ionization can fail is if there exists an $\sigma_{j},\phi_{j,0}(x,t)$ pair solving \eqref{eq:floquet} with $\Im \sigma_{j}=0$. Such a pair represents an eigenvalue/eigenvector pair associated to the Floquet Equation \eqref{eq:floquet} (with $N_{j}=1$) with $\phi_{j,0}(x,t)$ being $2\pi/\omega$-periodic in $t$ and square-integrable in space. If $\phi_{j,0}(x,t)$ were not square-integrable in space, then conservation of probability would be violated. Proving ionization then involves showing that nonzero solutions of \eqref{eq:floquet} do not exist.

Actually this dichotomy, either the system ionizes or there exists a square-integrable Floquet eigenstate, holds in more generality than \eqref{eq:resonanceExpansion}. It essentially rules out the existence of a singular spectrum of the Floquet Hamiltonian or solutions which are localized but not $2\pi/\omega$-periodic. This is proven to hold in all the cases we have investigated and we expect it to hold universally, c.f. \cite{galtbayer:floquetalternative,yajima:ACStarkExteriorScaling,yajima:ACStarkResonances}. We further expect that the first alternative, or $\Gamma_{j}>0$, will hold for essentially all $V_{0}(x,t)$ and $V_{1}(x,t)$. As shown below however, there are explicit examples where there are square-integrable solutions of \eqref{eq:floquet} and thus time-periodic bound states $\phi_{j,0}(x,t)$ (of period $2\pi/\omega$). In these cases the system never ionizes, unless of course $\psi(x,0)$ is orthogonal to $\phi_{j,0}(x,0)$.

We describe more explicitly our results for several systems. 

\begin{itemize}
\item We considered the widely studied \cite{susskind:multiphoton,MR1618647,Mercouris:deltafunctionstabilization} one dimensional system with $V_{0}(x)=-2 \delta(x)$ and dipole coupling to an oscillating electric field:
    \begin{equation}
      V_{1}(x,t)=\mathcal{E}(t)x = \sum_{j=1}^{\infty} \left[ \mathcal{E}_{j} e^{i j \omega t} + c.c.\right] x
    \end{equation}

  In \cite{us:dipoledelta} we prove for all field strength and all $\omega$ that \eqref{eq:resonanceExpansion} holds. We also show that if $\mathcal{E}(t)$ is a trigonometric polynomial, i.e. $\mathcal{E}_{j}=0$ for $j > M \geq 1$, then ionization occurs. For more general $\mathcal{E}(t)$ all we know is that (\ref{eq:resonanceExpansion}) holds implying that either the system ionizes or else it has a Floquet bound state. 
  \item Taking the same system with $V_{0}(x)=-2 \delta(x)$, but with $V_{1}(x,t)=\delta(x) \mathcal{E}(t)$, we showed that the system ionizes if $\mathcal{E}(t)$ is a trigonometric polynomial. In this case, we also proved that if $\mathcal{E}(t) = 2 a \lambda (\lambda-\cos(\omega t))/(1+\lambda^{2}-2\lambda \cos(\omega t))$, then there is a continuous set of values in the $(a, \lambda,\omega)$ parameter space for which ionization does not hold and a Floquet bound state exists, a strictly nonperturbative situation \cite{costin:cmpionize}. A similar situation occurs \cite{costin:jpaionize} when $V_{1}(x,t)=\mathcal{E} \sin(\omega t)[\delta(x-a)-\delta(x+a)]$ for a continuous set of $(\mathcal{E},a,\omega)$ even when $V_{0}(x)=0$. An example of a continuous $V_{0}(x)$ and $V_{1}(x,t)$ for which there are non-ionizing Floquet bound states can be constructed from breather modes of the nonlinear Schr\"odinger equation \cite{MR1758988}.

    These examples indicate that although uncommon, stabilization can occur when a resonance reaches the real line. It should be noted that in all these cases, the Floquet bound states are time-dependent and have many Fourier components in time. Even in the absence of bound states, it is possible for resonances $\sigma_{j}$ to move very close to the real line and survive for a long time. In fact we believe that such occurences correspond to what are usually termed Laser Induced Continuum States (LICS). LICS is a phenomenon in which, under the influence of a strong laser field, an atom with ground state $E_{0} < 0$ appears to form a bound state at energy $E_{0}+E_{s}+\omega > 0$ (with $E_{s}$ the Stark shift), which is located in the continuum \cite{lics:experimentalCoherentPopTransfer,lics:xenon,lics:experimentAndTheory,lics:continuumPopulationTransfer}. Our mode of analysis may form a means of studying  this phenomenon.
    
  \item Our last example is that of a Coulomb binding potential in $3$ dimensions: $V_{0}(x)=-Z e^{2}/\abs{x}$, and $V_{1}(x,t)$ periodic in time and compactly supported in space (i.e. $V_{1}(x,t)=0$ for $\abs{x} >R$). We show in this case that the dichotomy holds as well, i.e. either Floquet bound statesexist or ionization occurs. We prove the absence of Floquet bound states when $V_{1}(x,t)=\Omega(\abs{x}) \cos(\omega t)$ with $\Omega(\abs{x})>0$ for $\abs{x} \leq R$, and $\Omega(\abs{x})=0$ for $\abs{x} > R$ \cite{costin:coulomb}. This case is mathematically more difficult than the other cases due to the long range of the Coulomb potential which causes an accumulation of eigenvalues or resonances near zero energies. In this case, we have not yet proven \eqref{eq:resonanceExpansion}, though we are optimistic about proving a modified version of it (Eq. \eqref{eq:dispersiveExpansion} must altered to account for the long range potential).
\end{itemize}

Main steps of the proofs: The basic idea of our approach (which must be adapted to each case) is to relate the time asymptotics of $\psi(x,t)$ in the time domain with the analyticity properties of $\hat{\psi}(x,\sigma)$ in the energy domain. That is, we let $\bar{\psi}(x,p) = \int_{0}^{\infty} e^{-p t} \psi(x,t) dt$ be the Laplace transform of $\psi(x,t)$, and consider $\hat{\psi}(x,\sigma)$ with $\sigma=-i p$ so that $\sigma$ can be interpreted as a quasi-energy. The Schr\"odinger equation then becomes an inhomogeneous integral equation for $\hat{\psi}(x,\sigma)$. We then show that this equation is of a modified Fredholm type in a suitable Hilbert space. Application of the Fredholm Alternative yields the above mentioned dichotomy: either the homogeneous equation \eqref{eq:floquet} has a nonzero solution (leading to a Floquet bound state) or else $\hat{\psi}(x,\sigma)$ is analytic in the $\sigma$ variable on the inverse Laplace contour, except for a branch point near $\sigma=0$. The Riemann-Lebesgue lemma (stating that smoothness in $\sigma$ implies decay in $t$) implies ionization.

More quantitative results such as \eqref{eq:resonanceExpansion} are obtained by pushing the inverse Fourier contour into the lower half plane in $\sigma$. Resonances appear naturally as poles of $\hat{\psi}(x,\sigma)$ with $\Im \sigma < 0$ and bound states occur as poles with $\Im \sigma = 0$. Poles occur only where \eqref{eq:floquet} has a solution satisfying an outgoing wave condition. Standard  Fredholm theory shows solutions are analytic in both $\sigma$ and the field strength. The implicit function theorem therefore shows that the pole position is analytic (except for isolated singularities) in the field strength, which justifies perturbation theory. 

In the case of the Coulomb potential, the proof of compactness and analyticity is made difficult by the slow decay of Coulomb potential at $\infty$ which affects the decay rate of $\hat{\psi}(x,\sigma)$ as $\sigma \rightarrow \pm \infty$ on or close to the imaginary axis. Since the discrete spectrum has an accumulation point at zero, the Fredholm operator still has an essential singularity at $\sigma=0$ since $V_{1}(x,t)$ affects distant orbitals only minimally. Nonetheless, by appropriate choice of variables, it is possible to show boundedness as $\sigma=0$ is approached from the upper half-plane. The boundedness and continuity are reflected in the form of $\hat{\psi}(x,\sigma)$ as well: if indeed the associated homogeneous problem has only the zero solution, then ionization follows from Riemann-Lebesgue Lemma.

Perturbation theory in this formulation was carried out for the systems with a $\delta$-function binding potential and  $V_{1}(x,t)=\mathcal{E}(t) \delta(x)$ or $V_{1}(x,t)=\mathcal{E}(t)\cdot x$. In particular we showed that the leading order behavior is $\Gamma \sim (\mathcal{E}^{2})^{N}$, where $\mathcal{E}^{2}$ is the intensity of the radiation field and $N$ is the smallest integer for which $N \omega > -E_{0}$ with $E_{0}=-1$ the ground state energy of the unperturbed system. In the event that $N \omega = E_{0}$ for some $N$, then the perturbation expansion is done in powers of $\mathcal{E}$ rather than $\mathcal{E}^{2}$. The integer $N$ has the interpretation as the ``number of photons'' required for ionization. In the case when $\omega>1$ (corresponding to $N=1$) the lowest order term in the expansion reduces to Fermi's Golden Rule \cite{cohen:atomPhotonInteractions}. We expect this to be true generally.

To prove that ionization occurs, we need only prove that there are no localized solutions of \eqref{eq:floquet}. The method we have used in \cite{us:dipoledelta,costin:coulomb} is to solve \eqref{eq:floquet} for large $x$ and show that if the solution decays near $x = \infty$, then the solution must be discontinuous somewhere (e.g. at $x=0$). 

An example illustrating this method is given by $V_{0}(x)=-2 \delta(x)$ and $V_{1}(x,t)=\mathcal{E}(t) x$. In this case, we can avoid the use of WKB for large $x$ in lieu of an exact solution, and we use patching at $x=0$ instead of asymptotic matching. Elementary calculations show that if we ignore $V_{0}(x)=-2 \delta(x)$, \eqref{eq:floquet} has the family of solutions:
\newcommand{\fbasis}[2]{\chi_{#1,#2}}
\begin{subequations}
  \label{eq:freeSolutions}
  \begin{eqnarray}
    \fbasis{m}{\pm}(x,t) & =&  e^{\pm \lambda_m x} e^{- i m \omega t}
    e^{\pm \lambda_m c ( t )}\\
    \lambda_m & =&  -i \sqrt{\sigma_{j} + m \omega }\label{eq:lambdaDef} \\
    c ( t ) &=&  2 \sum_{n = 1}^{\infty} \left( \frac{\mathcal{E}_{n}}{( i n \omega )^2} e^{i n \omega t} + c.c. \right)
  \end{eqnarray}
\end{subequations}
If \eqref{eq:floquet} has a solution, to the left and right of $x=0$, we can expand it in terms of $\fbasis{m}{\pm}(x,t)$:
\begin{equation}
  \label{eq:solution of magnetic problem}
  \phi_{0}(x,t)=\left\{ \begin{array}{ll}
      \sum_{m < \Mo} \phi_{m}^L \fbasis{m}{+}(x,t), &  x \leq 0\\
      \sum_{m < \Mo}  \phi_{m}^R \fbasis{m}{-}(x,t), &  x \geq 0
    \end{array}  \right.
\end{equation}
The coefficients $\phi_{m}^{L,R}$ must be such that $\phi_{0}(x,t)$ is continuous at $x=0$ and the first derivative in $x$ must have an appropriate jump at $x=0$. In this expression, $\Mo$ is the smallest integer so that $\lambda_{m}$ is real and positive for all $m < \Mo$. The terms $\phi_{m}^{L}$ with $m \geq \Mo$ must be zero because any term with $m > \Mo$ would oscillate rather than decay for large negative $x$, thus making $\phi_{0}(x,t)$ non-integrable in space. Any term of the form $\fbasis{m}{-}(x,t)$ would grow exponentially as $x \rightarrow - \infty$, similarly making $\phi_{0}(x,t)$ non-integrable. The same argument applies to $x > 0$.

Physically, this means the following. If the resonant state $\phi_{j,0}(x,t)$ included terms with $m > \Mo$, it would be oscillatory near $x=\infty$. This would correspond to a state which radiates mass away without decreasing. This would violate the law of the conservation of probability, and is therefore impossible.

To prove ionization, we show that no function with the expansion \eqref{eq:solution of magnetic problem} can be continuous at $x=0$. In \cite{us:dipoledelta}, we use a Phragmen-Lindel\"of argument on the analytic continuation (in $t$) of $\phi_{0}(0,t)$. The argument shows that (using \eqref{eq:solution of magnetic problem} for $x \leq 0$) $\phi_{0}(0,t)$ must be analytic and bounded for $0 < \Re t < \pi, \Im t \geq 0$ and $\pi < \Re t < 2 \pi, \Im t \leq 0$. Similarly, we use the expansion for $x \geq 0$ to show boundedness for $0 < \Re t < \pi, \Im t \leq 0$ and $\pi < \Re t < 2 \pi, \Im t \geq 0$. The Phragmen-Lindelof theorem can then be used to show that $\phi_{0}(0,t)$ is bounded when $\Re t = 0,\pi,2\pi$. This implies that $\phi_{0}(0,t)$ is analytic and bounded, an impossible situation unless $\phi_{0}(0,t)=0$. This means that there exist no solutions to \eqref{eq:floquet}, thereby implying that all resonances in \eqref{eq:resonanceExpansion} are decaying.

For a Coulombic binding potential in 3 dimensions, $V_{0}(x)=-Z/\abs{x}$, and $V_{1}(x,t) = \Omega(\abs{x}) \cos(\omega t)$ (with $\Omega(\abs{x})=0$ for $\abs{x} > R$ and $\Omega(\abs{x}) > 0$ for $\abs{x} \leq R$ the analysis is accomplished similarly. Instead of trying to match solutions \eqref{eq:freeSolutions} at $x=0$ as in \eqref{eq:solution of magnetic problem}, we attempt to match Coulomb wave functions \cite[Chapter 14]{abramowitz:handbookmathfunctions} for $\abs{x} \geq R$ to time-periodic solutions of the full Schr\"odinger equation on the region $\abs{x} \leq R$. The proof of ionization again involves showing that no solution can be continuous and differentiable at the origin. Again, a crucial ingredient in the proof is the fact that the positive Fourier coefficients of $\phi_{j,0}(x,t)$ must be zero for $\abs{x} \geq R$, and required developing new asymptotic tools for the WKB analysis of infinite differential-difference systems.

\begin{acknowledgements}
  We thank R. Costin and A. Rokhlenko who participated in some of this work. We also thank K. Yajima, A. Soffer and M. Kiessling for helpful comments. Research supported by NSF Grants DMS-0100495, DMS-0406193, DMS-0600369, 
  DMS01-00490, 
  DMR 01-279-26 and AFOSR grant AF-FA9550-04. 
\end{acknowledgements}

\nocite{susskind:multiphoton}
\nocite{MR2163573}
\nocite{lics:xenon}
\nocite{MR1618647,schrader:anastab}
\nocite{lics:experimentAndTheory}
\nocite{PhysRevA.49.2117}
\nocite{geltman:unknown1}
\bibliography{../../stucchio}

\end{document}